\documentclass[12pt,preprint]{aastex}
\shorttitle{FUV cooling of Antlia SNR}
\shortauthors{Shinn et al.}

\newcommand{\cu}{ph s$^{-1}$ cm$^{-2}$ sr$^{-1}$ \AA$^{-1}$}
\newcommand{\lu}{ph s$^{-1}$ cm$^{-2}$ sr$^{-1}$}
\newcommand{\ncm}{cm$^{-3}$}

\newcommand{\Ncm}{cm$^{-2}$}
\newcommand{\kms}{km s$^{-1}$}

\newcommand{\Hi}{H {\small I}}
\newcommand{\Ha}{H${\alpha}$}

\newcommand{\Ciii}{C {\small III}}
\newcommand{\Civ}{C {\small IV}}

\newcommand{\Ni}{N {\small I}}

\newcommand{\Ovi}{O {\small VI}}
\newcommand{\Alii}{Al {\small II}}
\newcommand{\Siii}{Si {\small II}$^*$}

\newcommand{\ionCiii}{C$^{2+}$}
\newcommand{\ionCiv}{C$^{3+}$}

\newcommand{\dl}{$\lambda\lambda$}

\newcommand{\Eion}{$E_{ion}$}

\newcommand{\stsat}{\textit{STSAT--1}}

\begin{document}

\title{Far-ultraviolet Cooling Features of the Antlia Supernova Remnant}

\author{Jong-Ho Shinn\altaffilmark{1\dag}, Kyoung Wook Min\altaffilmark{1}, Ravi Sankrit\altaffilmark{3}, Kwang-Sun Ryu\altaffilmark{1}, Il-Joong Kim\altaffilmark{1}, Wonyong Han\altaffilmark{2}, Uk-Won Nam\altaffilmark{2}, Jang-Hyun Park\altaffilmark{2}, Jerry Edelstein\altaffilmark{3}, and Eric J. Korpela\altaffilmark{3}}

\email{jhshinn@kaist.edu}
\altaffiltext{1}{Korea Advanced Institute of Science and Technology (KAIST), 373-1 Guseong-dong, Yuseong-gu, Daejeon, 305-701, South Korea}
\altaffiltext{2}{Korea Astronomy and Space Science Institute (KASI), 61-1 Hwaam-dong Yuseong-gu Daejeon, 305-348, South Korea}
\altaffiltext{3}{Space Sciences Laboratory, University of California, Berkeley, CA 94720}
\altaffiltext{\dag}{Dept. of Physics and Astronomy, Seoul National University, San 56-1 Shillim9-dong, Gwanak-gu, Seoul, 151-747, South Korea}

\begin{abstract}
We present far-ultraviolet observations of the Antlia supernova remnant obtained with Far-ultraviolet IMaging Spectrograph (FIMS, also called SPEAR).
The strongest lines observed are \Civ{} \dl1548,1551 and \Ciii{} $\lambda$977.
The \Civ{} emission of this mixed-morphology supernova remnant shows a clumpy distribution, and the line intensity is nearly constant with radius.
The \Ciii{} $\lambda$977 line, though too weak to be mapped over the whole remnant, is shown to vary radially.
The line intensity peaks at about half the radius, and drops at the edge of the remnant.
Both the clumpy distribution of \Civ{} and the rise in the \Civ{} to \Ciii{} ratio towards the edge suggest that central emission is from evaporating cloudlets rather than thermal conduction in a more uniform, dense medium.
\end{abstract}

\keywords{ISM: individual (Antlia SNR) --- ISM: individual (Gum Nebula) --- ISM: individual (Vela SNR) --- (ISM:) supernova remnants --- ultraviolet: ISM}

\section{Introduction} \label{intro}
The Antlia supernova remnant (SNR) is a nearby ($d_A\simeq60-340$ pc), large ($\sim24\arcdeg$ in diameter), old ($t_A\sim1$ Myr) SNR recently discovered by \cite{McCullough(2002)ApJ_576_L41}, centered on  ($l=276\fdg52$, $b=+19\fdg05$).
It shows a center-filled X-ray morphology, which is partially enclosed by prominent \Ha{} and weak radio features (cf.~Fig.~\ref{antlia-mm}).
The remnant abuts the Gum Nebula to its southwest, and the boundary lies along the radio ridge seen in Fig.~\ref{antlia-mm}.
Based on the X-ray and radio morphologies, we classify the Antlia SNR as a mixed-morphology (MM) SNR \citep{Rho(1998)ApJ_503_L167}.
The X-ray is clearly center-filled, and though the radio shell is weak, there is no central radio emission.
Furthermore there is no known pulsar near the center of the Antlia SNR so it is unlikely to be pulsar-driven nebula.

To explain the physical characteristics of MM SNRs, in particular their center-filled X-ray morphologies, two models have been invoked.
Both focus on how the interior density, and thereby the emissivity may be increased.
In the first model \citep[hereafter ``cloudlet'' model]{White(1991)ApJ_373_543}, thermal evaporation of small clouds overrun by the supernova shock and engulfed by the hot gas bubble increases the density of the interior gas.
In the second model (hereafter ``dense medium'' model), applied to the SNR W44 by \cite{Cox(1999)ApJ_524_179}, thermal conduction in a dense medium alone increases the interior density of the remnant as heat is transported outward from the center.
These two models assume different distributions of the denser component that would show radiative-cooling
features as the remnant evolves.

Here, we report the detection of such cooling features in the far-ultraviolet (FUV) wavelengths from the Antlia SNR.
We observed the Antlia SNR with the Far-ultraviolet IMaging Spectrograph (FIMS, also called SPEAR), and detected several emission lines of which \Civ{} \dl1548,1551 and \Ciii{} $\lambda$977 are the strongest.
In the following sections we describe the observations, present the analysis and results, and discuss what we have learnt about MM SNRs.

\section{Observation and Data Reduction} \label{ond}
FIMS/SPEAR is a dual imaging spectrograph aboard the first Korean scientific satellite, \stsat, and is uniquely suited to observe the diffuse radiation in the FUV domain \citep{Edelstein(2006)ApJ_644_L153s}.
It has two spectral channels (S channel for $900-1150$ \AA{} and L channel for $1350-1750$ \AA), and their field-of-views are $4\fdg0\times4\farcm6$ and $7\fdg4\times4\farcm3$, respectively.
The spectral and imaging resolutions are, respectively, $\lambda/\Delta\lambda \sim 550$ and $\sim 5'$ on both channels.
The instrument, its on-orbit performance, and the data processing procedures are described in \cite{Edelstein(2006)ApJ_644_L159s}.

The \object{Antlia SNR} was observed during sky surveys between 2004 February 29 and April 19.
Some of the observations were performed with the shutter reduced to 10\% of the full slit length, and we also included them in the analysis.
Overall, the exposure time varies from $\sim1$ to $\sim30$ s.
We followed the general data processing procedures described in \cite{Edelstein(2006)ApJ_644_L159s} and then made a two-dimensional array for flexibility in handling the spectral and spatial information.
The spectral bin sizes were 0.5 \AA{} and 1.0 \AA{} for the S and L channels, respectively.
The spatial information was encoded by HEALPix pixelization \citep{Gorski(2005)ApJ_622_759}, and the pixel size was $\sim27\farcm5\times27\farcm5$.
A larger pixel size ($\sim55\farcm0\times55\farcm0$) was used in generating the emission-line map to enhance the signal intensity per pixel (cf.~next paragraph).
We adopted the effective area estimated in 2004 April for the S channel data since it is closer to the observation period than the 2003-November one (cf. \citealt{Edelstein(2006)ApJ_644_L159s}).

The emission-line map was produced through the following two steps: first, the spectrum of an individual pixel was fitted by a constant continuum plus a center-fixed Gaussian profile, with its width corresponding to the spectral resolution at the line center.
The fitting was carried out using a Bayesian method to estimate the signal amplitude in the presence of background assuming the Poisson statistics, as in \cite{Seon(2006)ApJ_644_L175s}.
This method avoids the over-subtraction, which often occurs when the continuum adjacent to the emission line is used as a model continuum.
Then, the emission-line map was smoothed with a Gaussian kernel, whose full-width-half-maximum is 2$\fdg$5, to enhance the features of any significance.

\section{Analysis and Results} \label{anr}
Following \cite{McCullough(2002)ApJ_576_L41}, we assumed the boundary of the Antlia SNR to lie roughly along a circle of 24\arcdeg{} in diameter (cf.~Fig.~\ref{antlia-mm}).
This boundary is defined mainly by the radio shell.
To the south, the \Ha{} emission also coincides with this shell.
There are X-ray emission features that lie outside this nominal boundary of the remnant.
This is not surprising since the direction towards the Antlia SNR is close to the Galactic plane and contains several active emission regions (including the Vela SNR).
In this analysis we do not analyze these regions, nor do we explore the possibility that they may be associated with the Antlia SNR.

Fig.~\ref{antlia-tsp} shows the total spectrum of the remnant extracted within the circle on Fig.~\ref{antlia-mm}, in CU($\equiv$ \cu; continuum unit).
In order to enhance the diffuse radiation relative to the radiation from point sources, we excluded the pixels of which wavelength-averaged intensity is greater than $10^5$ CU and $10^4$ CU for the S and L channels, respectively.
Several resonant emission lines are evident: \Ciii{} $\lambda$977, \Ni{} $\lambda$1135, \Siii{} $\lambda$1533, \Civ{} \dl1548,1551, and \Alii{} $\lambda1671$.
Si {\small II} $\lambda$1527, a doublet companion of \Siii{} $\lambda$1533 and a transition to the true ground state, was not seen in the spectrum, probably due to the resonant scattering by the interstellar warm ionized medium as in previous studies \citep{Shinn(2006)ApJ_644_L189s,Korpela(2006)ApJ_644_L163s,Kregenow(2006)ApJ_644_L167s}.

Table \ref{antlia-lu} lists the intensities of each identified emission line from the total spectrum in LU($\equiv$ \lu; line unit).
The observed intensities were extinction-corrected with $R_V=3.1$ Milky Way dust model \citep{Weingartner(2001)ApJ_548_296, Draine(2003)ARA&A_41_241}, adopting the hydrogen column density $N$(\Hi) $=3.0\times10^{20}$ \Ncm{} \cite{McCullough(2002)ApJ_576_L41} derived.
\Civ{} and \Ciii{} are the strongest lines and the ones that are associated with the gas in the SNR.
The other lines are typically too weak to be produced in shocks or thermal interfaces (e.g.~\citealt{Borkowski(1990)ApJ_355_501}) and probably arise in the nearby warm ionized medium due to photoionization.
Therefore we focus on these lines from ionized carbon, both of which are strong coolants of gas at around $10^5$ K (e.g.~\citealt{Cox(1971)ApJ_167_113}).
\Ovi{} was not detected in our spectrum, so we present the 90\% upper limit.
We note that the confidence interval of I(\Ovi)/I(\Ciii) from our data is consistent with the value found by \cite{Danforth(2003)ApJ_586_1179} for SNR 0057-7226, an MM SNR in the Small Magellanic Cloud, which they observed with the \emph{Far Ultraviolet Spectroscopic Explorer}.

Fig.~\ref{antlia-Civ} displays the emission-line map of \Civ{} \dl1548,1551.
The \Civ{} intensity is about $10000-20000$ LU over the remnant, and its diffuse background, measured over the ``BG'' box in Fig.~\ref{antlia-Civ}, is $3620\pm724$ LU.
It shows about $3\arcdeg-7\arcdeg$ scale features, which corresponds to $3-40$ pc adopting the inferred distance $d_A$; however, the minimum scale can be smaller since the resolution of the map is only about $2\fdg5$.
It generally shows a clumpy distribution all over the circle, and two intense-peak regions exist at both sides of the elongated hot gas seen in 3/4 keV map (see Fig.~\ref{antlia-mm}).
We could generate the emission-line map only for \Civ{}; the signal strength of \Ciii{} was not sufficient to generate a map.

Instead, we analyzed the radial variation of \Ciii{} intensity along with \Civ{} intensity.
Since both \ionCiii{} and \ionCiv{} originate from the same atom, but with different ionization energy (\Eion=24 eV for \ionCiii{} and \Eion=48 eV for \ionCiv), their line ratio is a useful diagnostic for the ionization state.
We chose subregions as in Fig.~\ref{antlia-Civ}, and estimated the intensity of \Ciii{} and \Civ{}.
The subregional spectra were extracted from \emph{two adjacent} subregions to get strong enough signals and simultaneously to see the average variation of the intensity profile.
This method is basically similar with the box-car smoothing of the ``true'' intensity profile.
The line intensities were estimated from each spectrum employing a $\chi^2$ minimization method \citep{Kriss(1994)inproc}.
The two \emph{upper} panels of Fig.~\ref{antlia-Cvar} show the subregional spectra around \Ciii{} and \Civ{}, respectively.
The two \emph{lower} panels of Fig.~\ref{antlia-Cvar} show the radial variations of their intensity and relative ratio, I(\Civ)/I(\Ciii), respectively.
\Civ{} intensity is nearly constant with radius while \Ciii{} intensity peaks at about half the radius, making their line ratio minimum near the \Ciii{} peak.

\section{Discussion}
The two models for MM SNRs, discussed in \S~\ref{intro}, use different distributions of the cold-and-dense gas component and therefore differ in their predictions of how the regions of radiative cooling
are distributed.  
The \Civ{} and \Ciii{} emission observed trace this cooling gas, and we now compare our observations to the model predictions.

\cite{Shelton(1999)ApJ_524_192} performed a 1-D numerical simulation on a SNR evolving in a dense ambient medium with a density gradient and thermal conduction.
Regardless of the thermal conduction mechanism, the $T=10^5$ K region, where \Civ{} emission is likely originate, is located almost at the same position---near the edge of the remnant.
Such an edge-concentration of \Civ{} was also predicted in the 1-D old-SNR model of \cite{Slavin(1992)ApJ_392_131}, which includes a similar thermal-conduction treatment.
Recently, \cite{Tilley(2006)MNRAS_371_1106} modeled the intensity of \Ovi{} \dl1032,1038, another FUV cooling line, for an MM SNR in 2-D under an anisotropic thermal conduction caused by magnetic fields.
Their result also shows an edge-concentration.
On the contrary, our \Civ{} map (Fig.~\ref{antlia-Civ}) shows a clumpy distribution over the remnant and \Civ{} peaks are seen at the central region.
This suggests that the ``dense medium'' model may not be suitable to explain the formation of the \object{Antlia SNR}.
The \Civ{} detection in the middle of the hot gas (region 3) in another MM SNR, the Lupus Loop, also supports this argument \citep{Shinn(2006)ApJ_644_L189s,Shinn(2007)phdth}.

There are no published predictions available for the \Civ{} emission from evaporating cloudlets.
However, we can qualitatively compare the expected distribution of \Civ{} emission from such a model with
our observations.
The clumpy distribution of \Civ{} emission can be explained, if small clouds were inhomogeneously scattered and engulfed by the hot bubble as in the ``cloudlet'' model.
The fact that there are numerous B stars in the vicinity of the Antlia SNR but no known O stars also supports the ``cloudlet'' model.
O-type stars may have sufficient photoionizing radiation and strong stellar wind to clear the molecular material in a region of $\sim15$ pc around them, while early B-type stars are not capable of this and may directly interact with molecular gas when they explode \citep{Chevalier(1999)ApJ_511_798}.
Thus, if the progenitor of the \object{Antlia SNR} were a B-type star, the remnant is likely to include interactions with the surrounding clouds.
In addition, the size of \Civ{} features is consistent with the ``cloudlet'' model.
\cite{Borkowski(1990)ApJ_355_501} computed the critical cloud radius of $\sim15$ pc, below which clouds are evaporating.
This scale is in agreement with the one seen in our \Civ{} map, $3-40$ pc, given the instrumental imaging resolution.

The \Civ{} intensity is nearly constant with radius along the direction measured, while the \Ciii{} intensity drops sharply at the edge (Fig.~\ref{antlia-Cvar}).
This suggests that the temperature is probably not decreasing near the edge as expected for the ``dense medium'' model \citep{Shelton(1999)ApJ_524_192}.
We note that the gas is probably not in ionization equilibrium, and that projection effects could influence the \Ciii{} to \Civ{} ratio.
Thus, the observed trend weakly supports the ``cloudlet'' model over the ``dense medium'' model.

In both ``cloudlet'' and ``dense medium'' models, the basic physical mechanism producing the FUV emitting gas is thermal conduction.
The line intensities expected from such a conduction front have been calculated by \cite{Slavin(1989)ApJ_346_718} and \cite{Borkowski(1990)ApJ_355_501}.
The former assumed a stationary interface and an oblique  magnetic field including the saturated thermal conduction, while the latter assumed an evolving interface and a perpendicular magnetic field neglecting the saturated thermal conduction.
Our observed line ratios between \Ciii{} and \Civ{}, and \Ovi{} and \Civ{} (Table \ref{antlia-lu}) are consistent with the evaporation phase of the \cite{Borkowski(1990)ApJ_355_501} models but not with the \cite{Slavin(1989)ApJ_346_718} models.

For a single evaporative interface at a pressure, $P/k_B=3750$ \ncm{} K, the \cite{Borkowski(1990)ApJ_355_501} models predict that the \Civ{} doublet intensity will be about 100 LU (see their Table 1).
The total intensity scales linearly with the pressure.
The interior pressure in a 1 Myr old remnant is about 10000 \ncm{} K \citep{Slavin(1992)ApJ_392_131}, and so the expected \Civ{} intensity is about 300 LU per interface.
We note that this is the intensity for an interface viewed face-on and that the observed brightness will be higher when viewed closer to edge-on, depending on the width of the emitting region (K. Borkowski, private communication).
The detected \Civ{} intensity (without the contribution of the background \Civ) is about 10,000 LU (Table \ref{antlia-lu}, and \S\ref{anr} above).
This is a factor about 30 higher than produced by a single interface viewed face-on.
Geometric enhancement of the emission due to the presence of many interfaces along the line of sight and due to viewing angles is necessary to account for the observed \Civ{} emission.
Each cloud contributes two interfaces if viewed close to face-on (one from the front and the other from the back) or a single interface with some amount of ``limb-brightening'' if viewed close to edge on.
Thus, 15 clouds could reproduce the observed \Civ{} emission, but the number could be much lower if the viewing angle is close to edge-on for one or more interfaces.
Clearly, if the pressure in the SNR were in fact higher than assumed for this calculation the geometric enhancement factor required would be correspondingly lower.

In astrophysical systems, FUV line emission can be produced in gas cooling from X-ray emitting temperatures and in shocked gas.
The first of these mechanisms is not likely for the emission observed from the Antlia SNR.
Cooling gas produces strong \Ovi{} emission, in general much stronger than \Civ{} \citep{Edgar(1986)ApJ_310_L27}, but which is not observed.
Shocks with velocities of about 110--130 \kms{} would produce line ratios similar to what we observed \citep{Hartigan(1987)ApJ_316_323}.
The shock model E110 of \cite{Hartigan(1987)ApJ_316_323} predicts that the \Civ{} intensity is 36000 LU when the pre-shock density ($N_0$) is 0.1 \ncm.
The line intensity is proportional to pre-shock density, and so the observed \Civ{} intensity will be produced by a 110 \kms{} shock running into gas with $N_0\sim0.03$ \ncm.
The driving pressure for such a shock, $P/k_B\sim36000$ \ncm{} K.
This is not far above the nominal pressure in a 1 Myr old remnant quoted above.
Alternatively, the pre-shock density could be as low as 0.01 \ncm{} and a modest geometric enhancement would be sufficient to produce the observed \Civ{} emission.
However, shock-cloud interaction regions exist at the boundary of the remnant.
If the \Civ{} emission detected is from the interior, then it is hard to invoke shocks as the exciting mechanism since the shock crossing times are much shorter \citep[e.g.][]{Cowie(1981)ApJ_247_908}.
If the the emitting regions are close to the boundary and their location near the middle of the remnant (Fig. 3) is a projection effect, then shocks cannot be ruled out as the mechanism for producing the FUV emission.  
However, as we have shown above, thermal conduction can produce the observed \Civ{} emission although some geometric enhancement is needed given the expected interior pressure in an old remnant such as the Antlia SNR.

\section{Conclusion}
We have presented and analyzed FUV observations of the Antlia SNR obtained with FIMS/SPEAR.
The Antlia SNR was identified as an MM SNR based on its X-ray and radio morphology, and its boundary was assumed as a 24$\arcdeg$-diameter circle at the center.
The clumpy \Civ{} map without an edge-concentration and the radial variation of I(\Civ)/I(\Ciii) within the remnant suggest that the ``cloudlet'' model, rather than the ``dense medium'' model, is plausible for the formation of the Antlia SNR's center-filled X-ray morphology.

The Antlia SNR is one of about 25 MM SNRs in the Galaxy, and perhaps the only one that can be studied in ultraviolet---all the others are suffering high extinction.
Hence, the results presented here are unique, showing that the ambient cloudlets may be an important component for energy exchange between MM SNRs and their surroundings.
However, it is clearly impossible to generalize from this one example to all objects in the class.
The ``dense medium'' model could be suitable in other situations where the ambient interstellar medium is different from that around the Antlia SNR.

We note that, at an angular resolution of $2\fdg5$, much of the sub-structure in the \Civ{} map is almost certainly hidden.
Thus, an FUV study of the Antlia SNR at much higher angular resolution (if and when that becomes possible) would allow the detailed structure of the radiatively cooling gas to be mapped and its relationship with the emission at other wavelengths to be characterized more precisely.

\acknowledgments
% short version
FIMS/SPEAR is a joint project of KASI \& KAIST (Korea) and U.C., Berkeley (USA), funded by the Korea MOST and NASA Grant NAG5-5355.
% long version
%FIMS/SPEAR is a joint project of Korea Advanced Institute  of Science and Technology, Korea Astronomy and Space Science Institute, and the University of California at Berkeley, funded by the Korea Ministry of Science and Technology and the U.S. National Aeronautics and Space Administration Grant NAG5-5355.
%Some of the results in this paper have been derived using the HEALPix package \citep{Gorski(2005)ApJ_622_759}.
The authors thank J. L. Jonas for the 2.3 GHz radio map toward the \object{Antlia SNR}, and also appreciate valuable comments from the referee and B.-C. Koo.

%\bibliographystyle{apj}
%\bibliography{jhshinn} 

\clearpage

\begin{figure}
%\center{
%\includegraphics[scale=0.4]{f1a.eps}\\
%\includegraphics[scale=0.4]{f1b.eps}
%}
%\epsscale{1}
%\plotone{f1a.eps}
%\plottwo{f1a.eps}{f1b.eps}
\caption{\textsf{The mixed-morphology of the Antlia SNR abutting on the Gum Nebula} The RASS 3/4 keV map \citep[\emph{top-panel},][]{Snowden(1995)ApJ_454_643} of the Antlia SNR is shown together with the contours of the Rhodes University/HartRAO 2.3 GHz radio continuum map \citep{Jonas(1998)MNRAS_297_977}. The radio features are seen as broken shells around the center-filled X-ray enhancement, especially at the bottom and right corners. The \Ha{} map \citep[\emph{bottom-panel},][]{Finkbeiner(2003)ApJS_146_407} shows a prominent arc feature along the faint radio shell. The Antlia SNR shows a simliar morphology with MM SNRs: the center-filled X-ray and shell-like radio morphology. The Antlia SNR abuts on the Gum Nebula at the lower-right corner. A circle of $24\arcdeg$ diameter, with its center at the core of the Antlia SNR, is shown on both maps. The coordinates are galactic. The units of colorbar are $10^{-6}$ counts s$^{-1}$ arcmin$^{-2}$ and Rayleighs, respectively.} \label{antlia-mm}
\end{figure}

\clearpage
\begin{figure}
\epsscale{0.6}
\plotone{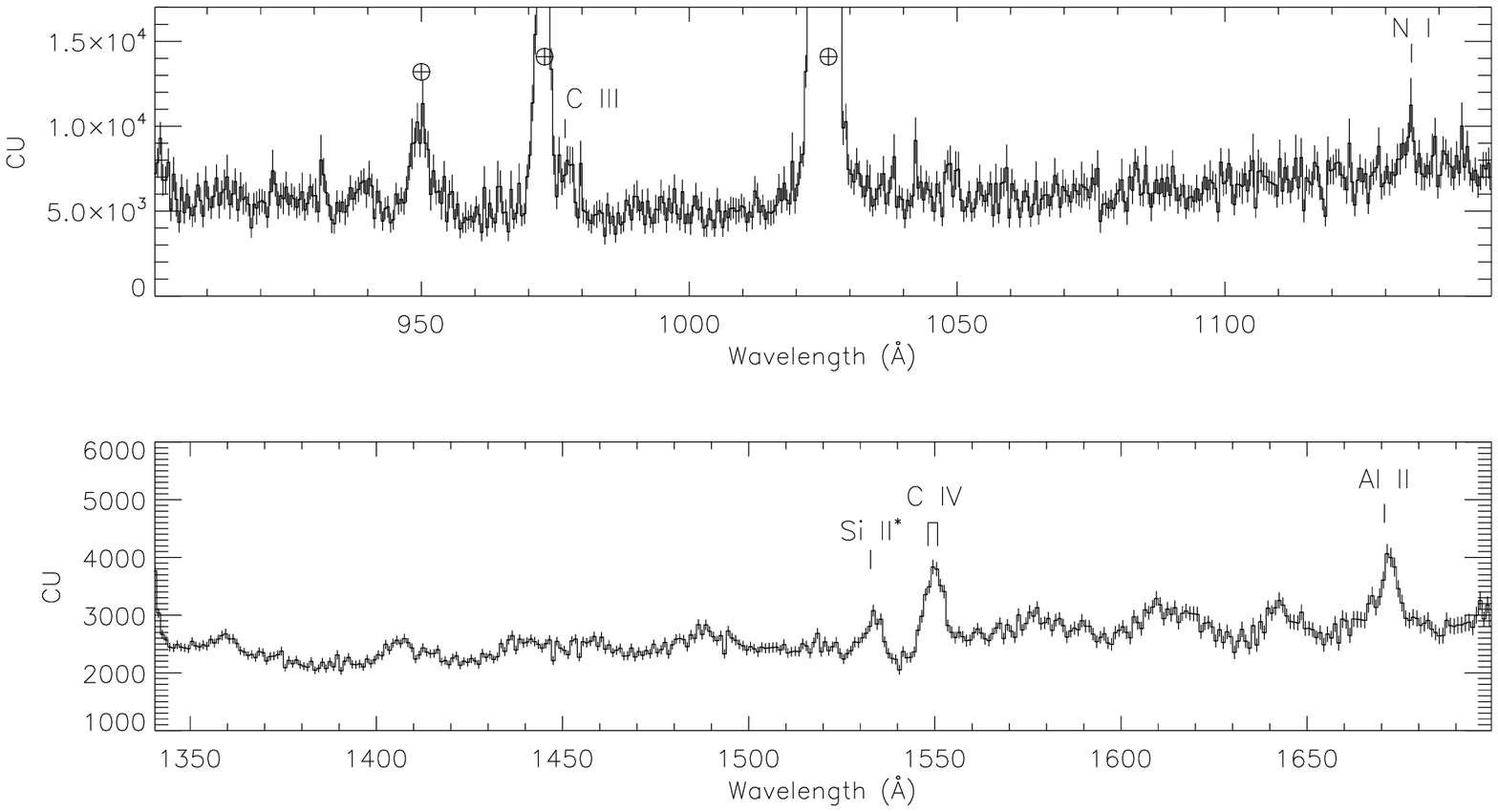}
\caption{\textsf{Emission lines identified from the total spectrum}  The total spectrum of each channel was extracted from the circle on Fig.~\ref{antlia-mm}.  They are binned in 0.5 \AA{} and 1.0 \AA{} for the Short and Long channels, respectively. \Ciii{} $\lambda$977, \Ni{} $\lambda$1135, \Siii{} $\lambda$1533, \Civ{} \dl1548,1551, and \Alii{} $\lambda1671$ are identified. `$\oplus$' indicates the geocoronal lines.} \label{antlia-tsp}
\end{figure}

\clearpage
\begin{figure}
%\includegraphics[scale=0.8]{f2.eps}
%\epsscale{0.55}
%\plotone{f3.eps}
\caption{\textsf{\Civ{} \dl1548,1551 emission-line map} The map was smoothed with a Gaussian kernel with $2\fdg5$ full-width-half-maximum. The unit of the colorbar is LU. The coordinates and the circles are the same as those in Fig.~\ref{antlia-mm}. The \Civ{} map generally shows a \emph{clumpy} distribution over the circle. The overlaid boxes inside the circle are the subregions where the spectra were extracted (cf.~Fig.~\ref{antlia-Cvar}). The size of one box is $7\fdg4\times2\fdg0$. The ``BG'' box is the region where the intensity of the diffuse background \Civ{} was estimated. It is $3620\pm724$ LU.} \label{antlia-Civ}
\end{figure}

\clearpage
\begin{figure}[p]
\centerline{
\includegraphics[angle=90,scale=0.29]{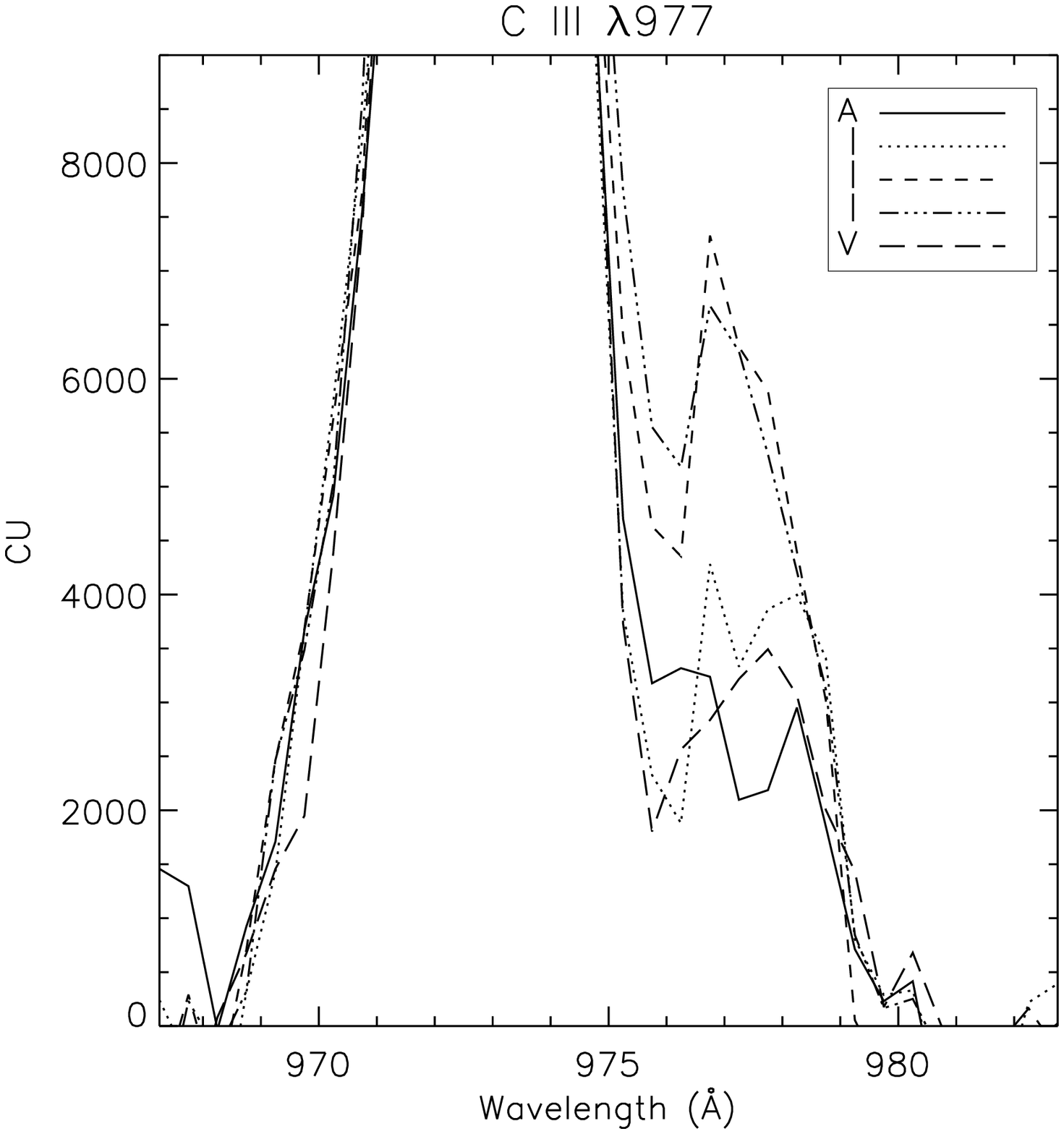}
\includegraphics[angle=90,scale=0.29]{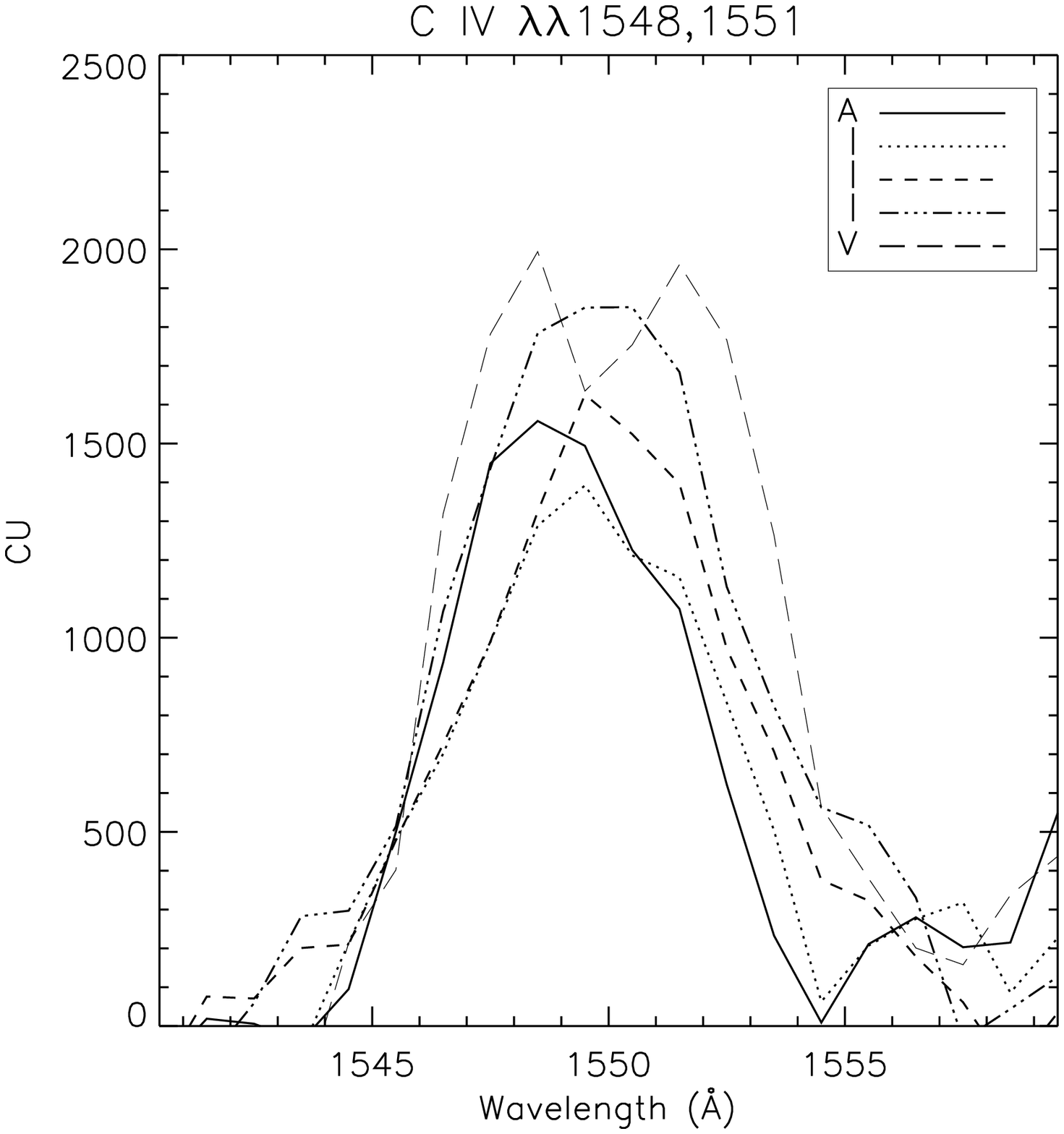}
}
\centerline{
\includegraphics[angle=90,scale=0.3]{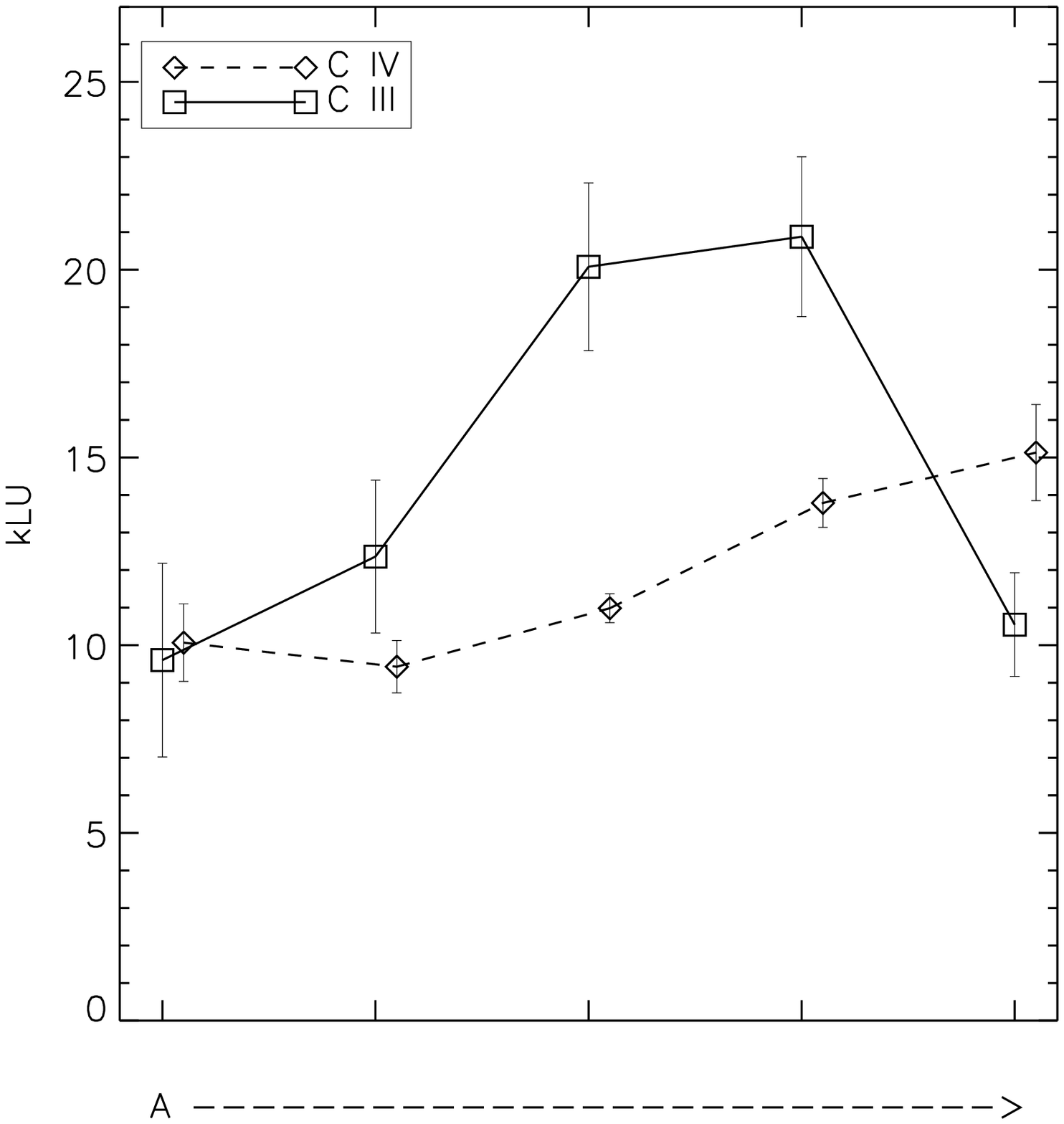}
\includegraphics[angle=90,scale=0.3]{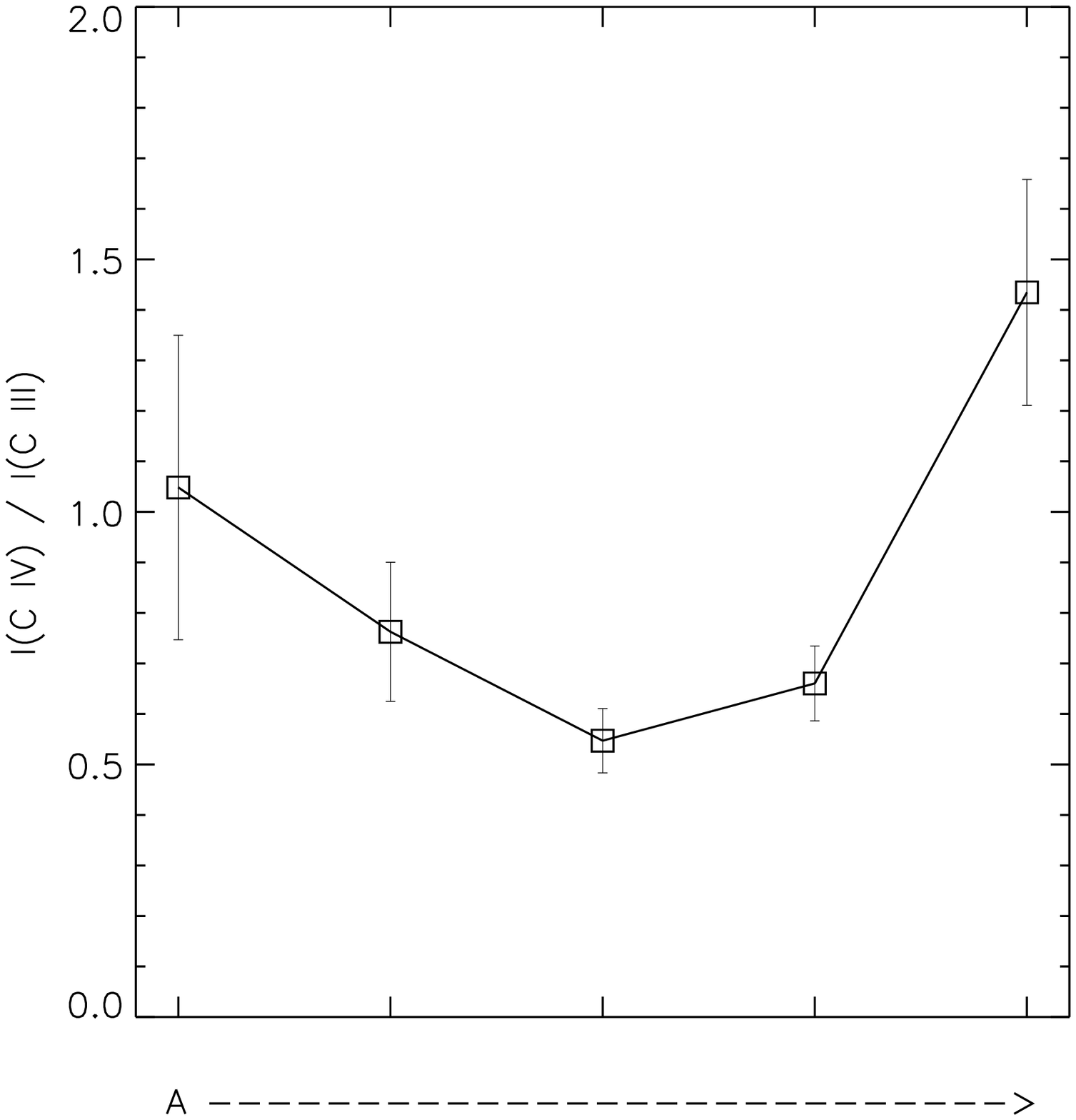}
}
\caption{\textsf{The radial variation of \Ciii{} $\lambda$977 and \Civ{} \dl1548,1551 intensity} The two \emph{upper} panels are the spectra around \Ciii{} $\lambda$977 and \Civ{} \dl1548,1551. They were extracted from \emph{two adjacent} subregions---note that the number of spectra is five, while the number of subregions is six (see Fig.~\ref{antlia-Civ} and text). The \emph{lower-left} panel plots the radial variations of their intensity. The \Civ{} mark is slightly shifted to the right in the \emph{lower-left} panel, to avoid the overlapping of the 1-$\sigma$ error bar. The \emph{lower-right} panel plots the radial variation of the ratio I(\Civ)/I(\Ciii). The \emph{rise} of I(\Civ)/I(\Ciii) near the edge suggests that the temperature is probably not decreasing outward.} \label{antlia-Cvar}
\end{figure}

\clearpage
\begin{deluxetable}{cccc}
\tablewidth{0pt}
\tablecaption{Intensity of each identified emission line from the total spectrum \label{antlia-lu}}
\tablehead{
&  & \colhead{Observed} & \colhead{\tablenotemark{\dag}Intrinsic}\\
\colhead{Emission Line} &  \colhead{Wavelength} & \colhead{Intensity} & \colhead{Intensity} \\
			&  \colhead{(\AA)} & \colhead{(LU)} & \colhead{(LU)}}
\startdata
%\Ciii{}      &977  &8165 $\pm$   1864 \\
%\Ovi{}      &1032, 1038  & $<$ 10463\tablenotemark{\dag}\\
%\Ni{}      &1135  &6252 $\pm$  1480 \\
%\Siii{}      &1533  &3582 $\pm$  403 \\
%\Civ{}      &1548, 1551  &9891  $\pm$ 526 \\
%\Alii{}      &1671  &6475  $\pm$ 623 

   \Ciii{}&977&8165$\pm$1864&15669$\pm$3578\\
    \Ovi{}&1032,1038&$<$10483\tablenotemark{\ddag}&$<$19267\tablenotemark{\ddag}\\
     \Ni{}&1134&6252$\pm$1480&10609$\pm$2511\\
   \Siii{}&1533&3582$\pm$403&5231$\pm$588\\
    \Civ{}&1548,1551&9891$\pm$526&14399$\pm$766\\
   \Alii{}&1671&6475$\pm$623&9306$\pm$895
\enddata

\tablenotetext{\dag}{Extinction-corrected intensity with $N$(\Hi)=$3.0\times10^{20}$ \Ncm{}. See text.}
\tablenotetext{\ddag}{90 \% upper limit.}

\end{deluxetable}

\end{document}